\documentstyle[12pt,axodraw]{article}
\begin{document}
\baselineskip 0.7cm

\newcommand{\gsim}{ \mathop{}_{\textstyle \sim}^{\textstyle >} }
\newcommand{\lsim}{ \mathop{}_{\textstyle \sim}^{\textstyle <} }
\newcommand{\vev}[1]{ \left\langle {#1} \right\rangle }
\newcommand{\EV}{ {\rm eV} }
\newcommand{\KEV}{ {\rm keV} }
\newcommand{\MEV}{ {\rm MeV} }
\newcommand{\GEV}{ {\rm GeV} }
\newcommand{\TEV}{ {\rm TeV} }
\def\tr{\mathop{\rm tr}\nolimits}
\def\Tr{\mathop{\rm Tr}\nolimits}
\def\Re{\mathop{\rm Re}\nolimits}
\def\Im{\mathop{\rm Im}\nolimits}
\setcounter{footnote}{1}

\begin{titlepage}

\begin{flushright}
UT-862\\
\end{flushright}

\vskip 2cm
\begin{center}
{\large \bf  Mass Generation for an Ultralight Axion}
\vskip 1.2cm
Yasunori Nomura$^1$, T.~Watari$^1$ and T.~Yanagida$^{1,2}$

\vskip 0.4cm

$^{1}$ {\it Department of Physics, University of Tokyo, \\
         Tokyo 113-0033, Japan}\\
$^{2}$ {\it Research Center for the Early Universe, University of Tokyo,\\
         Tokyo 113-0033, Japan}
\vskip 1.5cm

\abstract{If a global chiral symmetry is explicitly broken by anomalies
 in nonabelian gauge theories, a pseudo Nambu-Goldstone boson (axion)
 associated with a spontaneous breakdown of such a global symmetry
 acquires a mass through nonperturbative instanton effects.
We calculate the axion mass assuming a supersymmetric SU(2) gauge theory 
 and show that the axion obtains an extremely small mass when the SU(2)
 gauge symmetry is broken down at very high energy, say at the Planck
 scale.
We identify the axion with a hypothetical ultralight boson field
 proposed to account for a small but nonzero cosmological constant
 suggested from recent cosmological observations.}

\end{center}
\end{titlepage}

\renewcommand{\thefootnote}{\arabic{footnote}}
\setcounter{footnote}{0}

%
%
%
%

\section{Introduction}

A number of recent observations for the total (baryonic and dark) matter 
density from galaxy clusters suggest that it is significantly less than
the critical density \cite{Om_lt_1}.
However, most of the inflation models predict that the present universe
is spatially flat, namely the total energy density in the universe is
equal to the critical one \cite{Inflation}.
A nonzero cosmological constant seems to be the simplest candidate to
resolve this discrepancy without any contradiction to observations.
Furthermore, an introduction of a small cosmological constant provides
a natural solution to the so-called ``age crisis'' of the present
universe \cite{LCDM}.
It is also very encouraging that the recent studies on the Hubble
diagram for type Ia supernovae have supported the presence of the
nonzero cosmological constant \cite{SN_Ia}.
Therefore, it is quite reasonable to conclude that totality of the
present cosmological observations indicates a small but nonzero
cosmological constant.

However, a natural value of the cosmological constant 
$\Lambda_{\rm cos}^4$ is of order the Planck scale, $(M_{\rm pl} \simeq
2.4\times 10^{18}~\GEV)^4$, or at least of order the supersymmetry
(SUSY) breaking scale, $(m_{\rm SUSY} \simeq 1~\TEV)^4$, in
quantum field theories, since anything which contributes to vacuum
energy, such as a zero point energy of a field and a vacuum expectation
value of a field acts as a cosmological constant.
These are many orders of magnitude larger than the value 
$\Lambda_{\rm cos}^4 \simeq (3\times 10^{-3}~\EV)^4$ 
\cite{Om_lt_1, LCDM, SN_Ia} suggested from the cosmological
observations.
There has been found, so far, no satisfactory solution to this problem
\cite{CC_Review}.

At the present stage of understanding nature, thus, it seems very
interesting to assume that the energy density at the true vacuum of the
universe is exactly zero, due to some unknown mechanism, and to ask if
there is any mechanism that gives a small but nonzero cosmological
constant (effective vacuum energy) at the present epoch.
This assumption has led to an interesting hypothesis \cite{ULaxion1}
that an ultralight pseudo Nambu-Goldstone boson (axion) field ${\cal A}$ 
dominates the energy density of the present universe and its energy
density behaves like a cosmological constant.
When the expansion rate of the universe becomes smaller than the 
${\cal A}$ mass, the axion ${\cal A}$ will begin to oscillate around the 
true vacuum and the universe becomes eventually cold-axion dominated
with zero cosmological constant.\footnote{
There has been proposed an alternative mechanism \cite{Quint} to account 
for the small cosmological constant.}

In Refs.~\cite{ULaxion1, ULaxion2}, the axion field of mass $m_{\cal A}
\simeq 4\times 10^{-33}~\EV$ with a spontaneous breaking scale of a
global symmetry $F_{\cal A} \simeq 2\times 10^{18}~\GEV$ has been
considered to account for the small cosmological constant 
$\Lambda_{\rm cos}^4 \sim (3\times 10^{-3}~\EV)^4$ 
\cite{Om_lt_1, LCDM, SN_Ia}.
However, no convincing mechanism has been proposed to generate such an
extremely small mass for the axion.\footnote{
Ref.~\cite{ULaxion2} considers a neutrino mass $m_{\nu}$ as an explicit
breaking term of the global symmetry which induces a small mass for the
axion.
However, if one uses $m_{\nu} \simeq 5\times 10^{-2}~\EV$ suggested from 
the atmospheric neutrino oscillation \cite{atm_osc}, one gets too large
axion mass.}
In this paper, we propose a mechanism to generate the extremely small
mass for the axion ${\cal A}$.
Instantons of a broken gauge symmetry play a central role on producing
the axion mass in our model.

\section{A SUSY SU(2) Gauge Model}

The model is based on a SUSY SU(2) gauge theory with two doublet quarks
$Q^i$ $(i=1,2)$.
Here, we have suppressed the gauge index and $i$ denotes the flavor
index.
The extension to SU($N_c$) gauge theories $(N_c \geq 3)$ will be given
later.
The effective superpotential induced by the SU(2) gauge interaction is
\cite{eff_sup}
\begin{eqnarray}
  W_{\rm dyn} = \frac{2\Lambda^5}{\epsilon_{ij} Q^i Q^j},
\label{sup_dyn}
\end{eqnarray}
where $\Lambda$ is the renormalization-group invariant dynamical scale
of the SU(2) gauge theory.
The $\Lambda$ is given by
\begin{eqnarray}
  \Lambda = M_{\rm pl}\, {\rm e}^{-\frac{2\pi}{b\alpha(M_{\rm pl})}}.
\label{dyn_scale}
\end{eqnarray}
Here, the $b$ is the coefficient of one-loop renormalization-group
$\beta$ function, $b=5$ in the present model, and the 
$\alpha(M_{\rm pl}) \equiv g(M_{\rm pl})^2/4\pi$ the SU(2) gauge
coupling constant at the Planck scale $M_{\rm pl}$.
With this coupling $g$, the gauge kinetic term ${\cal L}_{\rm kin}$ is
defined as
\begin{eqnarray}
  {\cal L}_{\rm kin} = 
    \int d^2\theta \frac{1}{16g^2}{\cal W}^{a\alpha}{\cal W}^a_{\alpha}
    + {\rm h.c.},
\label{L_kin}
\end{eqnarray}
where ${\cal W}^{a\alpha}$ is the kinetic superfield of the SU(2) vector 
multiplet $(a=1-3)$ and $\alpha=1,2$ denotes the spinor index.
Notice that the effective superpotential $W_{\rm dyn}$ in
Eq.~(\ref{sup_dyn}) is obtained from nonperturbative instanton effects
\cite{eff_sup}.

For the region $Q^i \gg \Lambda$, the running gauge coupling constant
$\alpha(Q^i)$ is small and the K\"ahler potential for $Q^i$ is well
described by the minimal form
\begin{eqnarray}
  K \simeq Q_i^{\dagger} Q^i.
\label{K_pert}
\end{eqnarray}
With this K\"ahler potential and the dynamically induced superpotential
Eq.~(\ref{sup_dyn}), we easily see a runaway vacuum 
($Q^i \rightarrow \infty$) \cite{eff_sup}.
We introduce a gauge singlet chiral superfield $X$ to stabilize the
runaway vacuum in the perturbative regime ($\vev{Q^i} \gg \Lambda$).
We assume a simple superpotential for $X$ as
\begin{eqnarray}
  W_{\rm tree} = \frac{\kappa}{2} \epsilon_{ij} Q^i Q^j X
                 - V^2 X.
\label{sup_tree}
\end{eqnarray}
Then, the vacuum expectation values for the $Q^i$ and $X$ are
\begin{eqnarray}
  && \vev{\frac{1}{2} \epsilon_{ij} Q^i Q^j} = \frac{V^2}{\kappa},
\label{VEV_Q} \\
  && \vev{X} = \frac{\kappa\Lambda^5}{V^4},
\label{VEV_X}
\end{eqnarray}
where $Q^i$ and $X$ denote boson components of the corresponding
superfields.
Here, we have assumed that the scale $V$ is much larger than the
dynamical scale $\Lambda$, ${\it i.e.} V \gg \Lambda$, otherwise we get
$\vev{Q^i} \lsim \Lambda$ where the SU(2) gauge coupling is strong and
the perturbative approximation of the K\"ahler potential in
Eq.~(\ref{K_pert}) breaks down.

In this Higgs phase, the SU(2) gauge symmetry is spontaneously broken
down and all gauge multiplets become massive absorbing three massless
would-be Nambu-Goldstone (NG) multiplets $Q^i$'s.\footnote{
In this vacuum, we have an unbroken flavor SU(2)$_F$ symmetry.}
The remaining quark $Q^i$ also has a mass of order $\sqrt{\kappa}V$
together with the $X$ field.
Introduction of soft SUSY-breaking masses $m_Q^2$ and $m_X^2$ for the
scalar $Q^i$ and scalar $X$ fields does not induce any important changes 
in the mass spectrum and vacuum expectation values in Eqs.~(\ref{VEV_Q})
and (\ref{VEV_X}) as long as $V \gg m_Q^2, m_X^2$.
This is the case even for $m_Q^2, m_X^2 \gg \Lambda$.
For instance, the corrections to the vacuum expectation values
Eqs.~(\ref{VEV_Q}) and (\ref{VEV_X}) are of order $m_Q^2/V^2$ and
$m_X^2/V^2$, respectively, and the vacuum stays in the weak-coupling
regime.
This is a crucial point in our analysis, since we are interested in the
region $\Lambda^2 \ll m_{\rm SUSY}^2 \ll V^2 \simeq F_{\cal A}^2 \simeq
M_{\rm pl}^2$ as we shall see later, where $m_{\rm SUSY}^2$ denotes the
SUSY-breaking mass scale.\footnote{
For a region $V^2 \lsim \Lambda^2 \lsim m_{\rm SUSY}^2$, we have 
$\vev{Q^i} \lsim \Lambda$ and the perturbative description of the
K\"ahler potential Eq.~(\ref{K_pert}) breaks down.}

We now introduce a pseudo NG chiral superfield $\Phi(x,\theta)$ whose
boson components consist of the axion field ${\cal A}(x)$ and its scalar 
partner, saxion ${\cal B}(x)$.
We consider that the NG superfield is produced by a spontaneous
breakdown of some global symmetry at the Planck scale $F_{\cal A} \simeq
2\times 10^{18}~\GEV$, which is explicitly broken by the SU(2) gauge
anomaly.\footnote{
We neglect nonperturbative effects of gravitational interactions
\cite{NP_grav} on the axion mass, since we have not yet well understood
the quantum gravity.}
Then, we expect at low energies that the NG superfield $\Phi$ couples to
the SU(2) gauge multiplet through the anomaly as
\begin{eqnarray}
  {\cal L} = 
    \int d^2\theta \frac{1}{128\pi^2}\frac{\Phi}{F_{\cal A}}
    {\cal W}^{a\alpha}{\cal W}^a_{\alpha} + {\rm h.c.}
\label{coup_NG}
\end{eqnarray}
Here, the axion ${\cal A}(x)$ and saxion ${\cal B}(x)$ fields are an
imaginary and a real part of the complex boson component of
$\Phi(x,\theta)$, respectively.

Together with superpotentials Eqs.~(\ref{sup_dyn}), (\ref{sup_tree}) and 
(\ref{coup_NG}), we now obtain a low-energy effective
superpotential\footnote{
We redefine the kinetic term ${\cal L}_{\rm kin}$ as ${\cal L}_{\rm kin}
= \int d^2\theta (G/4){\cal W}^{a\alpha}{\cal W}^a_{\alpha} +
{\rm h.c.}$ with $G = 1/4g^2 + \Phi/(32\pi^2F_{\cal A})$,
combining Eq.~(\ref{L_kin}) and Eq.~(\ref{coup_NG}).
Then, the $W_{\rm dyn}$ is given by $W_{\rm dyn} = 2M_{\rm pl}^5
\exp(-32\pi^2 G)/\epsilon_{ij} Q^i Q^j$.
This leads to Eq.~(\ref{sup_tot}).}
\begin{eqnarray}
  W_{\rm eff} = \frac{2\Lambda^5}{\epsilon_{ij} Q^i Q^j}
                {\rm e}^{-\Phi/F_{\cal A}}
              + \frac{\kappa}{2} \epsilon_{ij} Q^i Q^j X - V^2 X.
\label{sup_tot}
\end{eqnarray}
Then, integration of the $X$ and the $Q^i$ yields
\begin{eqnarray}
  W_{\rm eff} = \frac{\kappa\Lambda^5}{V^2}{\rm e}^{-\Phi/F_{\cal A}}.
\label{sup_eff}
\end{eqnarray}
With this superpotential, the ${\cal B}$ has a runaway behavior and the
effective gauge coupling constant defined by $g_{\rm eff} = 
(1/g^2 + \vev{\cal B}/(8\pi^2 F_{\cal A}))^{-1/2}$ goes to zero.
Thus, the ${\cal A}$ is massless at the limit ${\cal B} \rightarrow
\infty$, since the nonperturbative effects which would generate the
axion mass vanishes when $g_{\rm eff} \rightarrow 0$.

The K\"ahler potential for the $\Phi$ superfield should respect the
global symmetry $\Phi \rightarrow \Phi + i\delta$.
Then, we have the following form of K\"ahler potential:
\begin{eqnarray}
  K = \frac{1}{2}(\Phi + \Phi^{\dagger})^2 
      + \frac{c}{F_{\cal A}}(\Phi + \Phi^{\dagger})^3
      + \cdots.
\end{eqnarray}
Notice that the linear term has been eliminated by a shift of the $\Phi$ 
field of order $F_{\cal A}$, which is accompanied by a slight shift of
the gauge coupling constant, $g$, through Eq.~(\ref{coup_NG}).
If we consider the SUSY breaking in a hidden sector in supergravity
\cite{hidden}, the soft SUSY-breaking mass 
$m_{\cal B}^2 \simeq m_{\rm SUSY}^2$ for the ${\cal B}$ field arises
with $m_{\rm SUSY}$ being the gravitino mass.
The soft mass for the ${\cal B}$ field, $m_{\cal B}^2$, stops the
runaway behavior of the ${\cal B}$ at
\begin{eqnarray}
  \vev{\cal B} \simeq \frac{\kappa^3\Lambda^{10}}
    {m_{\cal B}^2 F_{\cal A}^3 V^6} 
    \left( \frac{V^2}{\kappa}-F_{\cal A}^2 \right).
\label{VEV_B}
\end{eqnarray}
Since we consider the region $\Lambda^2 \ll m_{\cal B}^2 \ll
F_{\cal A}^2 \simeq V^2 \simeq M_{\rm pl}^2$,
$\vev{\cal B} \simeq \Lambda^{10}/m_{\cal B}^2 M_{\rm pl}^7 \ll 
F_{\cal A}$ and we set $\exp(-{\cal B}/F_{\cal A}) \simeq 1$, hereafter.
Even then ($g_{\rm eff} \simeq g \neq 0$), however, the ${\cal A}$ field
remains massless.
This is because there is an anomaly-free global symmetry that is a
linear combination of the global symmetry 
$\Phi \rightarrow \Phi + i\delta$ and an $R$-symmetry 
$X \rightarrow \exp(2i\delta)X$, which is broken down by the
vacuum expectation value $\vev{X} \simeq
(\kappa\Lambda^5/V^4)$ in Eq.~(\ref{VEV_X}).\footnote{
The massless field has a small admixture of the phase of the $X$ field
as ${\cal A} + (\vev{X}/F_{\cal A})\varphi_X \simeq
{\cal A} + (\kappa\Lambda^5/F_{\cal A}V^4)\varphi_X$, where 
$X = \vev{X}\exp(i\varphi_X/|\vev{X}|)$.}
Thus, we have to introduce an explicit breaking of the $R$-symmetry in
order to generate an axion mass.

\section{Induced Potential for the Axion}

We now calculate the axion mass, introducing the $R$-breaking gaugino
mass $m_{\tilde{g}}$ of the SU(2) gauge theory.\footnote{
Introduction of the gaugino mass $m_{\tilde{g}}$ shifts the vacuum
expectation value of the ${\cal B}$ to $\vev{\cal B} \simeq 
(\kappa \Lambda^5 / m_{\cal B}^2 F_{\cal A} V^2) 
(32\pi^2 m_{\tilde{g}} / g^2)$ from that in Eq.~(\ref{VEV_B}).
Even in this case, however, the $\vev{\cal B}$ is still sufficiently
small, $\vev{\cal B} \ll F_{\cal A}$, to fix 
$\exp(-{\cal B}/F_{\cal A}) \simeq 1$.}$^{,}$\footnote{
A SUSY-breaking trilinear boson coupling, $(\kappa/2) A \epsilon_{ij}
Q^i Q^j X$, may also play a similar role to the gaugino mass
$m_{\tilde{g}}$ on inducing the axion mass, where $Q^i$ and $X$ are
boson components of the corresponding superfields.}
The effect of the gaugino mass can be incorporated by promoting the
gauge coupling to the superfield $S$ as
\begin{eqnarray}
  {\cal L}_{\rm kin} = 
    \int d^2\theta \frac{1}{4}S{\cal W}^{a\alpha}{\cal W}^a_{\alpha}
    + {\rm h.c.},
\end{eqnarray}
and regarding its $F$-component as the gaugino mass \cite{spurion},
\begin{eqnarray}
  S = \frac{1}{4g^2} - \frac{i\Theta}{32\pi^2} 
      - \frac{m_{\tilde{g}}}{2g^2}\theta^2.
\end{eqnarray}
Here, the $\Theta$ is the vacuum angle.
Then, the dynamical scale $\Lambda$ defined by Eq.~(\ref{dyn_scale}) is
also promoted to the superfield as
$\Lambda^b = M_{\rm pl}^b \exp(-32\pi^2S)$.
Substituting this into the superpotential Eq.~(\ref{sup_eff}) and taking 
the $F$-component of the superpotential, we get a
mass term for ${\cal A}$ as
\begin{eqnarray}
  {\cal L}_{{\cal A}\,{\rm mass}} &\simeq& 
      \frac{16\pi^2m_{\tilde{g}}}{g^2} \frac{\kappa\Lambda^5}{V^2}
      {\rm e}^{-i\frac{\cal A}{F_{\cal A}}} + {\rm h.c.} \nonumber\\
  &=& \left| \frac{8\pi m_{\tilde{g}}}{\alpha}
      \frac{\kappa\Lambda^5}{V^2} \right|
      \cos\left( \frac{\cal A}{F_{\cal A}} + \Theta^{\prime} \right),
\label{axion_pot}
\end{eqnarray}
where $\Theta^{\prime} = \Theta + \arg(V^2) - \arg(\kappa) -
\arg(m_{\tilde{g}})$.
Then, the axion mass is given by
\begin{eqnarray}
  m_{\cal A}^2 \simeq 
    \left| \frac{8\pi\kappa}{\alpha}
    \frac{m_{\tilde{g}}\Lambda^5}{F_{\cal A}^2V^2} \right|,
\label{axion_mass}
\end{eqnarray}
which is extremely small in the region $\Lambda \ll m_{\tilde{g}} \ll
F_{\cal A} \simeq V \simeq M_{\rm pl}$.\footnote{
Strictly speaking, the axion field is a linear combination of ${\cal A}$
and phases of $X$ and $Q$, since the mass matrix for the axion and the
phase fields is given by
\begin{eqnarray}
  {\cal L} \simeq \frac{1}{2}
        \pmatrix{{\cal A}\, \varphi_X\, \varphi_Q  \cr}
        \pmatrix{
          \frac{8\pi\kappa}{\alpha}
              \frac{m_{\tilde{g}}\Lambda^5}{F_{\cal A}^2V^2} &
            \kappa^2 \frac{\Lambda^5}{F_{\cal A}V^2} &
            \kappa^{7/2} \frac{\Lambda^{10}}{F_{\cal A}V^7}  \cr
          \kappa^2 \frac{\Lambda^5}{F_{\cal A}V^2} &
            \kappa V^2 &
            \kappa^{5/2} \frac{\Lambda^5}{V^3}  \cr
          \kappa^{7/2} \frac{\Lambda^{10}}{F_{\cal A}V^7} &
            \kappa^{5/2} \frac{\Lambda^5}{V^3} &
            \kappa V^2  \cr}
        \pmatrix{
          {\cal A} \cr  \varphi_X \cr  \varphi_Q  \cr}. 
\end{eqnarray}
Here, $\varphi_X$ and $\varphi_Q$ are defined as 
$X = \vev{X}\exp(i\varphi_X/|\vev{X}|)$ and 
$\epsilon_{ij}Q^iQ^j/2 = \vev{\epsilon_{ij}Q^iQ^j/2}
\exp(i\varphi_Q/\sqrt{|\vev{\epsilon_{ij}Q^iQ^j/2}|})$,
respectively.
However, the mixing between ${\cal A}$ and the phases $\varphi_{X,Q}$
are very small.
Thus, the axion is dominantly the ${\cal A}$ field and its mass is given 
by Eq.~(\ref{axion_mass}).}
This result can be also derived explicitly from the instanton
calculation which is given in the Appendix.

It is very intriguing to identify the above axion with the hypothetical
ultralight boson field proposed \cite{ULaxion1, ULaxion2} to explain the 
small cosmological constant $\Lambda_{\rm cos}^4 \simeq (3\times
10^{-3}~\EV)^4$ suggested from the recent observations.
This scenario requires $m_{\cal A} \simeq 4\times 10^{-33}~\EV$.
For $F_{\cal A} \simeq V \simeq 2\times 10^{18}~\GEV$, the extremely
small axion mass $m_{\cal A}$ is obtained with a moderate value of the
dynamical scale $\Lambda \simeq 10^{-3}~\GEV$\footnote{
If the gauge symmetry was not broken, one would need an extremely small
dynamical scale $\Lambda \simeq 10^{-3}~\EV$ to obtain such a small
axion mass $m_{\cal A} \simeq 4\times 10^{-33}~\EV$.}
which corresponds to the SU(2) gauge coupling constant 
$\alpha(M_{\rm pl}) \simeq 1/38$ at the Planck scale.
Here, we have assumed the Yukawa coupling $\kappa \simeq 1$ and the
gaugino mass $m_{\tilde{g}} \simeq 1~\TEV$.
It is very encouraging that it is fairly closed to the standard-model
gauge coupling constants at the Planck scale.

We have presented a model that gives naturally an extremely small mass
for the ultralight axion through the instanton effects of a broken SU(2) 
gauge theory.\footnote{
One may use instantons of the electroweak SU(2)$_L$ gauge theory in the
SUSY standard model as suggested in Ref.~\cite{ULaxion1}.
In this case, the axion acquires a mass from the K\"ahler potential
\cite{SUSY_axi}.
Under the presence of $(B+L)$-violating dimension-five operators, we
have estimated the axion mass as $m_{\cal A}^2 \lsim \left| 
(\mu m_{\rm SUSY}^3/F_{\cal A}^2) \exp(-2\pi/\alpha_2(M_{\rm pl}))
\right| \simeq (10^{-36}~\EV)^2$ for $\alpha_2(M_{\rm pl}) \simeq 1/24$
and $F_{\cal A} \simeq M_{\rm pl}$.
Here, $\mu \simeq 1~\TEV$ is the SUSY-invariant Higgs mass.
This axion mass is much smaller than the required value.
However, if there are matter multiplets which increase the SU(2)$_L$
gauge coupling constant, $\alpha_2(M_{\rm pl})$, at the Planck scale,
one may obtain the desired value for the axion mass.}
The extension of the above model to the case of SU($N_c$) with $N_c-1$
pairs of quarks and antiquarks is straightforward.
In this case, the axion mass is given by
\begin{eqnarray}
  m_{\cal A}^2 \simeq 
    \left| \frac{8\pi\kappa^{N_c-1}}{\alpha}
    \frac{m_{\tilde{g}}\Lambda^{2N_c+1}}{F_{\cal A}^2V^{2N_c-2}} \right|.
\end{eqnarray}
Even in these models, however, the required value for the gauge coupling
constant $\alpha(M_{\rm pl})$ is the same as in the case of the SU(2).

\section*{Acknowledgments}

We are grateful for discussions with K.-I.~Izawa, M.~Kawasaki and
K.~Kurosawa.
Y.N. thanks the Japan Society for the Promotion of Science for financial
support.
This work was partially supported by ``Priority Area: Supersymmetry and
Unified Theory of Elementary Particles (\# 707)'' (T.Y.).

\newpage

\section*{Appendix}

In this Appendix, we calculate the axion potential Eq.~(\ref{axion_pot})
in terms of the component-field instanton calculation.
The relevant parts of the Euclidean action $S_{\rm E}$ is given by
\begin{eqnarray}
  - S_{\rm E} &=& - \int d^4x_{\rm E}
  \Biggl( 
    \frac{({\cal B}+i{\cal A})}{32\pi^2 F_{\cal A}}
    \frac{1}{2}(F^a_{\mu\nu}F^a_{\mu\nu}-F^a_{\mu\nu}\tilde{F}^a_{\mu\nu})
\nonumber\\
  && \qquad\qquad
     + \frac{({\cal B}-i{\cal A})}{32\pi^2 F_{\cal A}}
     \frac{1}{2}(F^a_{\mu\nu}F^a_{\mu\nu}+F^a_{\mu\nu}\tilde{F}^a_{\mu\nu}) 
\nonumber\\
  && \qquad\qquad
     + {\cal D}_{\mu}Q_i^*{\cal D}_{\mu}Q^i
     - \sqrt{2}i(Q_i^* \lambda^a \frac{\tau^a}{2} \psi^i)
     - \frac{m_{\tilde{g}}}{2g^2} \lambda^a \lambda^a
  \Biggr),
\label{Euclid}
\end{eqnarray}
where $\tau^a$ is the Pauli matrices, $\lambda^a$ the SU(2) gaugino and
$Q^i(x,\theta) = Q^i(x) + \sqrt{2}\theta\psi^i(x)$.
We define instanton and anti-instanton configurations as those
satisfying $F^a_{\mu\nu} = \tilde{F}^a_{\mu\nu}$ and 
$F^a_{\mu\nu} = -\tilde{F}^a_{\mu\nu}$, respectively.
Then, one anti-instanton effect generates the superpotential
Eq.~(\ref{sup_dyn}) \cite{eff_sup}.

The zero-mode configuration on one anti-instanton background is given by 
\cite{Instanton}
\begin{eqnarray}
  Q^*_{li}(x) &=&
    \frac{(-i\bar{\sigma}_{\mu})_{li}(x-x_0)_{\mu}}
    {((x-x_0)^2+\rho^2)^{1/2}}
    \left( \frac{V^*}{\sqrt{\kappa^*}} \right), \\
  \psi^{li}_{\alpha}(x) &=& 
    \frac{2\sqrt{2}\,\delta^l_{\alpha}\rho^2}{((x-x_0)^2+\rho^2)^{3/2}}
    \left( \frac{V}{\sqrt{\kappa}} \right) \bar{\zeta}_0^i, \\
  \lambda^a_{\alpha}(x) &=&
    \frac{-8(\tau^a)_{\alpha}^{\beta}(-i\sigma_{\mu})_{\beta\dot{\alpha}}
    (x-x_0)_{\mu}\rho^2}{((x-x_0)^2+\rho^2)^2} \bar{\beta}^{\dot{\alpha}}
    + \frac{-8i(\tau^a)_{\alpha}^{\beta}\rho^2}
    {((x-x_0)^2+\rho^2)^2} \theta_{0\beta},
\end{eqnarray}
where $l,m = 1,2$ are the SU(2) gauge indices, and $\alpha,\beta=1,2$
and $\dot{\alpha}=1,2$ denote the left-handed and right-handed spinor
indices, respectively.
Here, $x_0$ and $\rho$ represent the center and the size of the
instanton, respectively, and $\theta_0, \bar{\beta}, \bar{\zeta}_0$ are 
anticommuting variables.
Substituting this classical configuration into the action
Eq.~(\ref{Euclid}) and expanding functional measure around the
configuration, we obtain the instanton measure
\begin{eqnarray}
  d\mu &=& d^4x_0\, d\rho^2\, d^2\theta_0\, d^2\bar{\beta}\, 
    d^2\bar{\zeta}_0
    \frac{\kappa}{4 \rho^2 V^2} \Lambda^5 
\nonumber\\ 
  && \exp \Biggl( -\int d^4x 
    \left( \frac{({\cal B}+i{\cal A})(x)}{F_{\cal A}}
    \frac{6\rho^4}{\pi^2((x-x_0)^2+\rho^2)^4} \right)
\nonumber\\ 
  && \qquad -4\pi^2\rho^2 \left|\frac{V}{\sqrt{\kappa}}\right|^2 
    + 16i\pi^2\rho^2 \left|\frac{V}{\sqrt{\kappa}}\right|^2
      \bar{\beta}\bar{\zeta}_0 
    + \frac{m_{\tilde{g}}}{2g^2} 32\pi^2 \theta_0^2 \Biggr).
\end{eqnarray}
Here, we have suppressed measures and interactions among fluctuation
modes.\footnote{
Strictly speaking, $\rho$ is an unstable mode and not a collective
coordinate, since the instanton configuration is not a stationary point
of the Euclidean action for the spontaneously broken gauge theory.}

\begin{figure}
\begin{center}
\begin{picture}(600,300)(70,0)

 \CArc(160,150)(30,0,360)           \Text(160,150)[]{$AI$}
 \ArrowArc(108,120)(52,60,145)       \PhotonArc(108,120)(52,60,145){3}{7}
 \ArrowArcn(108,180)(52,300,215)
 \Vertex(65,150){2}
 \DashArrowLine(65,150)(45,150){4}    \Text(45,140)[]{$\vev{Q^*}$}
                       \Line(42,153)(48,147) \Line(42,147)(48,153)
 \ArrowArcn(212,120)(52,120,35)     \PhotonArc(212,120)(52,35,120){3}{7}
 \ArrowArc(212,180)(52,240,325)
 \Vertex(255,150){2}
 \DashArrowLine(255,150)(275,150){4}\Text(275,140)[]{$\vev{Q^*}$}
                       \Line(272,153)(278,147) \Line(272,147)(278,153)
 \ArrowArc(190,98)(52,150,235)      \PhotonArc(190,98)(52,150,235){3}{7}
 \ArrowArcn(130,98)(52,30,305)      \PhotonArc(130,98)(52,305,30){3}{7}
 \Vertex(160,56){2}
 \DashArrowLine(160,21)(160,56){4}
                          \Line(157,18)(163,24) \Line(163,18)(157,24)
                          \Text(170,21)[l]{$-\frac{m_{\tilde{g}}}{2g^2}$}
 \DashArrowLine(90,220)(139,171){4} 
          \Text(90,230)[b]{$\frac{{\cal B}+i{\cal A}}{F_{\cal A}}$}
 \DashArrowLine(230,220)(181,171){4} 
          \Text(230,230)[b]{$\frac{{\cal B}+i{\cal A}}{F_{\cal A}}$}
 \DashArrowLine(146,210)(152,180){4}\DashArrowLine(162,195)(161,182){4} 
 \DashArrowLine(173,208)(165,180){4}\DashArrowLine(185,195)(174,178){4} 

\SetWidth{1.5}
 \Line(290,150)(310,150)
 \Line(300,140)(300,160)
\SetWidth{0.5}

 \CArc(440,150)(30,0,360)           \Text(440,150)[]{$I$}
 \ArrowArcn(388,120)(52,145,60)     \PhotonArc(388,120)(52,60,145){3}{7}
 \ArrowArc(388,180)(52,215,300)
 \Vertex(345,150){2}
 \DashArrowLine(325,150)(345,150){4} \Text(325,140)[]{$\vev{Q}$}
                       \Line(322,153)(328,147) \Line(322,147)(328,153)
 \ArrowArc(492,120)(52,35,120)      \PhotonArc(492,120)(52,35,120){3}{7}
 \ArrowArcn(492,180)(52,325,240)
 \Vertex(535,150){2}
 \DashArrowLine(555,150)(535,150){4} \Text(555,140)[]{$\vev{Q}$}
                       \Line(552,153)(558,147) \Line(552,147)(558,153)
 \ArrowArcn(470,98)(52,235,150)     \PhotonArc(470,98)(52,150,235){3}{7}
 \ArrowArc(410,98)(52,305,30)       \PhotonArc(410,98)(52,305,30){3}{7}
 \Vertex(440,56){2}
 \DashArrowLine(440,56)(440,21){4}
                        \Line(437,18)(443,24) \Line(443,18)(437,24)
                        \Text(450,21)[l]{$-\frac{m^*_{\tilde{g}}}{2g^2}$}
 \DashArrowLine(419,171)(370,220){4}
          \Text(370,230)[b]{$\frac{{\cal B}-i{\cal A}}{F_{\cal A}}$}
 \DashArrowLine(461,171)(510,220){4} 
          \Text(510,230)[b]{$\frac{{\cal B}-i{\cal A}}{F_{\cal A}}$}
 \DashArrowLine(448,180)(454,210){4}\DashArrowLine(439,182)(438,195){4} 
 \DashArrowLine(435,180)(427,208){4}\DashArrowLine(426,178)(415,195){4} 

\end{picture}
\end{center}
\caption{The diagrams which generate the axion potential.
The left one represents the anti-instanton contribution giving 
$\exp(-({\cal B}+i{\cal A})/F_{\cal A})$ potential, while the right one
 represents the instanton contribution giving 
$\exp(-({\cal B}-i{\cal A})/F_{\cal A})$ potential.}
\label{fig_pot}
\end{figure}
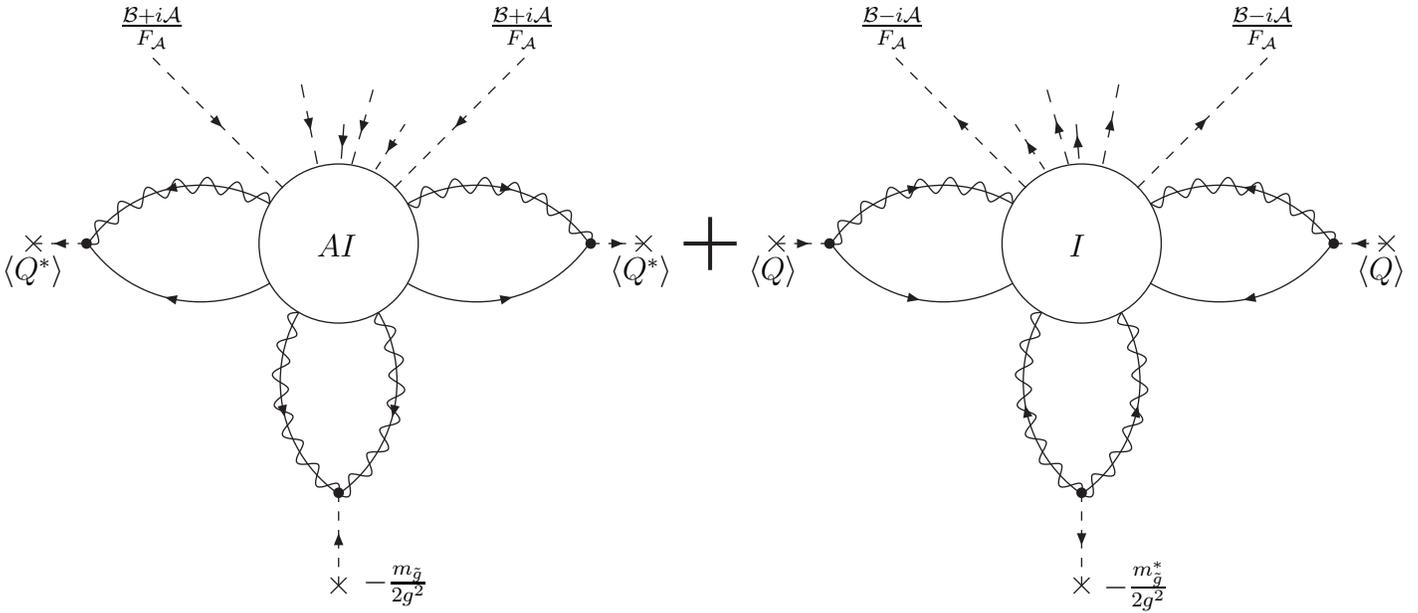

Integration of fermionic coordinates, $\theta_0, \bar{\beta}$ and
$\bar{\zeta}_0$, yields
\begin{eqnarray}
  d\mu &=& d^4x_0\, d\rho^2
    16\pi^4 \rho^2 \frac{16\pi^2 m_{\tilde{g}}}{g^2} 
    \frac{V^{*2}\Lambda^5}{\kappa^*} 
    \exp \left(-4\pi^2\rho^2 \left|\frac{V}{\sqrt{\kappa}}\right|^2\right)
\nonumber\\ 
  && \times \exp \left( -\int d^4x 
    \left( \frac{({\cal B}+i{\cal A})(x)}{F_{\cal A}}
    \frac{6\rho^4}{\pi^2((x-x_0)^2+\rho^2)^4} \right) \right).
\label{instanton_measure}
\end{eqnarray}
The factor $m_{\tilde{g}}/2g^2$ comes from an insertion of the gaugino
mass, $V^{*2}/\kappa^* = \vev{Q^*}^2$ from matter-gaugino
vertices, and $\Lambda^5 = M_{\rm pl}^5\exp(-2\pi/\alpha(M_{\rm pl}))$
from an anti-instanton, which can be easily read off from a diagrammatic
expression in Fig.~\ref{fig_pot}.
Notice that the $R$-breaking term,
$-(m_{\tilde{g}}/2g^2) \lambda^a \lambda^a$, is necessary in order to
obtain a nonzero axion potential.

Inspecting Eq.~(\ref{instanton_measure}), we see that only small-scale
instanton contributes to the axion potential, since integration of the
instanton size, $\rho$, is cut off at $\rho \simeq
|\sqrt{\kappa}/2\pi V|$ due to the exponential factor 
$\exp(-4\pi^2\rho^2 |V/\sqrt{\kappa}|^2)$.
Then, the smearing factor $d^4x\, 6\rho^4/\pi^2((x-x_0)^2+\rho^2)^4$ is 
well approximated by the delta function $\delta^4(x-x_0)$ as long as we
consider the effective potential at low energies.
Then, we can perform the $d\rho^2$ integral and obtain 
\begin{eqnarray}
  d\mu &=& d^4x_0
    \frac{16\pi^2 m_{\tilde{g}}}{g^2} 
    \frac{\kappa \Lambda^5}{V^2} 
    \exp\left(-\frac{({\cal B}+i{\cal A})(x_0)}{F_{\cal A}}\right).
\end{eqnarray}
Recalling that the ${\cal B}$ is fixed at 
$\vev{\cal B} \ll F_{\cal A}$ by the soft SUSY-breaking mass 
$m_{\cal B}^2$ (see the text), we determine the axion potential induced
by one anti-instanton as
\begin{eqnarray}
  {\cal L}_{{\cal A},\, {\rm AI}} \simeq 
    \frac{16\pi^2m_{\tilde{g}}}{g^2} \frac{\kappa\Lambda^5}{V^2}
    {\rm e}^{-i\frac{\cal A}{F_{\cal A}}}.
\label{axion_AI}
\end{eqnarray}
Similar calculation shows that one instanton induces the axion potential 
\begin{eqnarray}
  {\cal L}_{{\cal A},\, {\rm I}} \simeq 
    \frac{16\pi^2m_{\tilde{g}}^*}{g^2} \frac{\kappa^*\Lambda^{*5}}{V^{*2}}
    {\rm e}^{i\frac{\cal A}{F_{\cal A}}},
\label{axion_I}
\end{eqnarray}
through the right diagram in Fig.~\ref{fig_pot}.
We obtain the axion mass term given in Eq.~(\ref{axion_pot}) by summing
up both anti-instanton and instanton induced potentials,
Eqs.~(\ref{axion_AI}) and (\ref{axion_I}).

\newpage
%
%
%
\newcommand{\Journal}[4]{{\sl #1} {\bf #2} {(#3)} {#4}}
\newcommand{\PL}{\sl Phys. Lett.}
\newcommand{\PR}{\sl Phys. Rev.}
\newcommand{\PRL}{\sl Phys. Rev. Lett.}
\newcommand{\NP}{\sl Nucl. Phys.}
\newcommand{\ZP}{\sl Z. Phys.}
\newcommand{\PTP}{\sl Prog. Theor. Phys.}
\newcommand{\NC}{\sl Nuovo Cimento}
\newcommand{\MPL}{\sl Mod. Phys. Lett.}
\newcommand{\PRep}{\sl Phys. Rep.}

%
\newpage
\end{document}